\documentclass[aps,showpacs,preprintnumbers,amsmath, amssymb]{revtex4}

\usepackage{float}
\usepackage{graphics,epsfig}
\usepackage{graphicx}
\usepackage{dcolumn}
\usepackage{bm}

\begin{document}
\baselineskip=0.8 cm

\title{{\bf No short hair behaviors of ultra-compact stars }}
\author{Yan Peng$^{1}$\footnote{yanpengphy@163.com}}
\affiliation{\\$^{1}$ School of Mathematical Sciences, Qufu Normal University, Qufu, Shandong 273165, China}

\vspace*{0.2cm}
\begin{abstract}
\baselineskip=0.6 cm
\begin{center}
{\bf Abstract}
\end{center}

In the black hole spacetime, a no short hair theorem was proved,
which states that the  effective radius of black hole hairs must
extend beyond the null circular orbit. In the present paper,
in the horizonless gravity, we find a similar no short hair
behavior that the effective radius of matter fields must also
extend beyond the null circular orbit of ultra-compact stars.

\end{abstract}

\maketitle
\newpage
\vspace*{0.2cm}

\section{Introduction}

One of the most interesting predictions of general relativity is
the existence of null circular orbits outside compact astrophysical objects,
such as black holes and horizonless ultra-compact stars \cite{c1,c2}.
These null circular geodesics provide the way that massless
particles (photons, gravitons) can orbit a central
compact objects, which gives important information about the
highly curved spacetime. At present, the null circular orbits have been widely studied
\cite{ad1}-\cite{r2}.

Recently, it was proved that the null circular orbit
can be used to describe the distribution of exterior matter fields
outside black holes. In spherically symmetric hairy black hole spacetimes,
it was found that the effective radius of hairs must
extend beyond the null circular orbit \cite{s1,s2}.
The authors in \cite{s1} have proposed a nice heuristic physical picture
that the null circular orbit divides matter fields
into two parts (fields below the orbit tending to be swallowed
by black holes and fields above the orbit tending to be radiated away to infinity)
and the self-interaction between these two parts bind
together fields in different regions leading to
the existence of black hole hairs.
And it was further conjectured that the region above the null circular orbit
contains at least half of the total hair mass \cite{s2,s3,s4,s5}.

The main goal of this paper is to study properties of
horizonless ultra-compact spherical stars with null circular orbits.
We shall analytically prove a no short hair theorem that
the effective radius of matter fields must
extend beyond the null circular orbit.

\section{Investigations on effective radius of matter fields}

We consider static horizonless stars with null circular orbits.
In Schwarzschild coordinates, these ultra-compact spherically
symmetric objects take the following form \cite{s1,s2}
\begin{eqnarray}\label{AdSBH}
ds^{2}&=&-g(r)e^{-2\chi(r)}dt^{2}+\frac{dr^{2}}{g(r)}+r^{2}(d\theta^2+sin^{2}\theta d\phi^{2}).
\end{eqnarray}
The metric functions $\chi(r)$ and $g(r)$ only
depend on the Schwarzschild radial coordinate r.
Asymptotical flatness of the spacetime at the infinity
requires that $g(r\rightarrow \infty)=1$ and $\chi(r\rightarrow \infty)=0$.
Near the origin, regularity conditions of the gravity are \cite{b1,b2}
\begin{eqnarray}\label{AdSBH}
g(r\rightarrow 0)=1+O(r^2)~~~~~~and~~~~~~\chi(0)<\infty.
\end{eqnarray}

We take $\rho=-T^{t}_{t}$, $p=T^{r}_{r}$ and
$p_{T}=T^{\theta}_{\theta}=T^{\phi}_{\phi}$,
where $\rho$, $p$ and $p_{T}$ are interpreted as
the energy density, the radial pressure and
the tangential pressure respectively \cite{s2}.
Einstein differential equations $G^{\mu}_{\nu}=8\pi T^{\mu}_{\nu}$ yield
the metric equations
\begin{eqnarray}\label{BHg}
g'=-8\pi r \rho+\frac{1-g}{r},
\end{eqnarray}
\begin{eqnarray}\label{BHg}
\chi'=\frac{-4\pi r (\rho+p)}{g}.
\end{eqnarray}

We label $m(r)$ as the gravitational mass contained within a sphere of radial radius r.
It can be expressed by the integral relation
\begin{eqnarray}\label{AdSBH}
m(r)=\int_{0}^{r}4\pi r'^{2}\rho(r')dr'.
\end{eqnarray}
Considering relations (3) and (5), one
can express the metric function $g(r)$ in
the form \cite{b2}
\begin{eqnarray}\label{BHg}
g=1-\frac{2m(r)}{r}.
\end{eqnarray}

Following approaches in \cite{s2}, we obtain equations of null circular orbits.
As the metric (1) is independent of the time $t$ and angular coordinates $\phi$,
the geodesic trajectories are characterized by conserved energy
E and conserved angular momentum L. And the null circular orbits are
determined by an effective potential
\begin{eqnarray}\label{BHg}
V_{r}=(1-e^{2\chi})E^{2}+g\frac{L^2}{r^2}
\end{eqnarray}
with the characteristic relations
\begin{eqnarray}\label{BHg}
V_{r}=E^{2}~~~~~~and~~~~~~V_{r}'=0.
\end{eqnarray}

Substituting Einstein equations (3) and (4) into
(7) and (8), we deduce the null circular orbit equation
\cite{b1}
\begin{eqnarray}\label{BHg}
N(r_{\gamma})=3g(r_{\gamma})-1-8\pi (r_{\gamma})^2p(r_{\gamma})=0,
\end{eqnarray}
where we have introduced a new function $N(r)=3g(r)-1-8\pi (r)^2p(r)$.
The discrete roots of (9) correspond to the radii of the null circular orbit.

With the relation (2) and the regular condition $p(0)<\infty$, one obtains
\begin{eqnarray}\label{BHg}
N(0)=2.
\end{eqnarray}

We label $r_{\gamma}^{in}$ as the radius of the innermost null circular orbit.
That is to say $r_{\gamma}^{in}$ is the smallest positive root of $N(r)=0$.
Also considering relation (10), one deduces that
\begin{eqnarray}\label{BHg}
N(r)\geqslant0~~~~for~~~~r\in [0,r_{\gamma}^{in}].
\end{eqnarray}

The conservation equation $T^{\mu}_{\nu;\mu}=0$ has only one nontrivial component
\begin{eqnarray}\label{BHg}
T^{\mu}_{r;\mu}=0.
\end{eqnarray}

Substituting equations (3) and (4) into (12),
one gets equation of the pressure
\begin{eqnarray}\label{BHg}
p'(r)=\frac{1}{2rg}[(3g-1-8\pi r^2p)(\rho+p)+2gT-8gp]
\end{eqnarray}
with $T=-\rho+p+2p_{T}$ standing for the trace of the energy momentum tensor.

With a new pressure function $P(r)=r^4p$,
the equation (13) can be expressed as
\begin{eqnarray}\label{BHg}
P'(r)=\frac{r}{2g}[N(\rho+p)+2gT],
\end{eqnarray}
where $N=3g-1-8\pi r^2p$.

The energy condition usually plays a
central role in determining the spacetime geometry.
We assume the dominant energy condition
\begin{eqnarray}\label{BHg}
\rho\geqslant |p|,~|p_{T}|\geqslant 0,
\end{eqnarray}
which means that the energy density $\rho$ is non-negative and
bounds the pressures \cite{s1,s2}. We also assume the
non-negative trace of the energy momentum tensor
expressed as
\begin{eqnarray}\label{BHg}
T=-\rho+p+2p_{\tau}\geqslant 0.
\end{eqnarray}
In this work, we take the non-negative trace conditions \cite{b1,b2,b3,b4}.
A well known example for such horizonless configurations
is the gravitating Einstein-Yang-Mills solitons,
which are characterized by the identity $T=0$ \cite{T}.
In contrast, the usual non-positive trace condition was
imposed in the black hole spacetime \cite{s1,s2}.

From (11) and (14-16), one deduces that
\begin{eqnarray}\label{BHg}
P'(r)\geqslant 0~~~~for~~~~r\in [0,r_{\gamma}^{in}].
\end{eqnarray}

We impose the condition that $\rho$ goes to zero
faster than $r^{-4}$ \cite{s1,s2}. Also considering (15)
and $P(r)=r^4p(r)$, there is the asymptotical behavior
\begin{eqnarray}\label{BHg}
P(r\rightarrow \infty)=0.
\end{eqnarray}

Near the origin, the pressure function $P(r)$ has the asymptotical behavior
\begin{eqnarray}\label{BHg}
P(r\rightarrow 0)=0.
\end{eqnarray}

Relations (18) and (19) imply that $|P(r)|$ must have a local maximum value
at some extremum point $r_{0}$. We can define $r_{m}=r_{0}$ as the effective radii
of matter fields \cite{s1,s2}. According to (17), $P(r)$ is an increasing function of r before it
reaches the innermost null circular orbit.
So the effective radii have a lower bound
\begin{eqnarray}\label{BHg}
r_{m}\geqslant r_{\gamma}^{in},
\end{eqnarray}
which is the same as the no short hair theorem of black holes \cite{s1,s2}.

\section{Conclusions}

We investigated distributions of matter fields
in the background of horizonless ultra-compact spherically symmetric stars.
We assumed the dominant energy condition and the non-negative trace condition.
We defined an effective matter field radius at an extremum point
where the pressure function $|P(r)|$ possesses a local maximum value.
Using analytical methods, we obtained a lower bound
on the effective matter field radius expressed as
$r_{m}\geqslant r_{\gamma}^{in}$ with
$r_{m}$ as the effective matter field radius
and $r_{\gamma}^{in}$ corresponding to the innermost null circular orbit radius.
So we found a no short hair behavior that the
effective matter field radius must extend beyond the
innermost null circular orbit.

\begin{acknowledgments}

This work was supported by the Shandong Provincial
Natural Science Foundation of China under Grant
No. ZR2018QA008. This work was also supported by
a grant from Qufu Normal University of China under
Grant No. xkjjc201906.

\end{acknowledgments}

\end{document}